\documentclass[pre,superscriptaddress,showkeys,showpacs,twocolumn]{revtex4-1}
\usepackage{graphicx}
\usepackage{latexsym}
\usepackage{amsmath}
\begin {document}
\title {Diffusive behavior of a greedy traveling salesman}
\author{Adam Lipowski}
\affiliation{Faculty of Physics, Adam Mickiewicz University, Pozna\'{n}, Poland}
\author{Dorota Lipowska}
\affiliation{Faculty of Modern Languages and Literature, Adam Mickiewicz University, Pozna\'{n}, Poland}
\begin {abstract}
Using  Monte Carlo simulations we examine the diffusive properties of the greedy algorithm in the $d$-dimensional traveling salesman problem. 
Our results show that for $d=3$ and 4 the average squared distance from the origin $\langle r^2\rangle$ is proportional to the number of steps $t$. In the $d=2$ case such a scaling is modified with some logarithmic corrections, which might suggest that $d=2$ is the critical dimension of the problem.
The distribution of lengths also shows marked differences between $d=2$ and $d>2$ versions. A simple strategy adopted by the salesman might resemble strategies chosen by some foraging and hunting animals, for which anomalous diffusive behavior has recently been reported and interpreted in terms of L\'evy flights. Our results suggest that broad and L\'evy-like distributions  in such systems might appear due to dimension-dependent properties of a search space.
\end{abstract}

\maketitle
\section{Introduction}
Although the classical random walk problem is rather well understood by now~\cite{weiss}, some of its modifications, related,  for example, to disordered systems, are less clear ~\cite{bouchaud}. Departure from the classical random walk manifests often as an
anomalous diffusion that appears whenever the typical square displacement of a tracer particle  $R^2(t)$ does not obey  the normal diffusion law $R^2(t)\sim t^{\alpha}$ with $\alpha=1$.  Deviations include numerous examples of subdiffusion ($\alpha<1$) such as, for example, photoconductivity of amorphous materials \cite{pfister}, diffusion in convective rolls \cite{pomeau}, or a random walk on the percolation cluster \cite{havlin}. On the other hand, there is much rarer superdiffusive ($\alpha>1$) behavior, of which the Richardson diffusion in turbulent fluids is a classical example \cite{rich}.
Some macromolecules or biological systems also exhibit anomalous dynamics with superdiffusive behavior, as reported, e.g., for micelles \cite{Ott} or living cell migrations \cite{Dieterich}.

Recently, a number of studies have indicated that foraging or hunting strategies of some animals  also show superdiffusive behavior and examples include ants \cite{schlesinger}, fruit flies  \cite{cole}, or wandering albatrosses \cite{stanley}. The prevailing understanding is that while searching for food these animals move in steps of varying length with fat power-law distribution consistent with the L\'evy flights.  However, some other works show that this hypothesis should be taken with care. Indeed, re-analyzing the data for the wandering albatrosses, Edwards~{\it et~al.}\ concluded that the previously suggested L\'evy-flight scenario appeared due to a spurious inclusion of very long flights \cite{edwards}. It has also been claimed that the idea of animals using stochastic but essentially blind search strategies does not seem to be viable since it neglects the role of animals' intelligence and experience in guiding them \cite{travis}.  Nevertheless, some more extensive recent studies for 14 species of open-ocean fish of prey seem to support the L\'evy-flight scenario \cite{humphries}.

In the present article we show that a simple search strategy, i.e.: go to the nearest (intelligence) and not yet visited (experience) place, leads to an interesting diffusive behavior. In particular, a normal diffusion that takes place for $d=3$ and 4 systems is modified by some logarithmic corrections in the $d=2$ case, which suggests that this is the critical dimension. The distribution of distance does not have fat power-law tails, however, when the search exploits all available places, much broader and possibly L\'evy-like distributions can be found.

\section{The model and the numerical method}
\begin{figure}
\includegraphics[width=9cm]{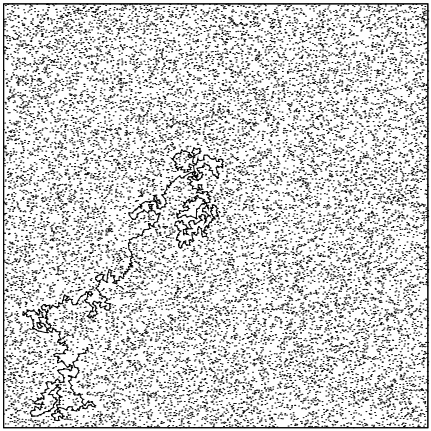}  \vspace{-0.7cm} 
\caption{ 
An exemplary configuration of the path of the greedy traveling salesman for $N=2\times 10^4$ cities (dots) and duration $t=10^3$.
\label{conf}}
\end{figure}

We consider a traveling salesman \cite{tsp} that has to visit cities which are uniformly distributed in the \mbox{$d$-di}mensional space of linear size $L$.  We assume a unit density of cities, i.e., $N/L^d=1$, where $N$ is the number of cities.
The salesman is not interested in finding the minimal tour but visits cities according to the following greedy algorithm: in each step, from the last visited city  it moves to the nearest city which has not been visited yet. 
For most simulations, the starting point is the city placed in the center of the space, i.e., of the Cartesian coordinates $(L/2,L/2,...,L/2)$.
The quantity of main interest is the average square distance $\langle r^2\rangle$ from the starting point and especially its asymptotic form after many steps~$t$. An example of a typical trajectory in the two-dimensional case is shown in Fig.~\ref{conf}. 
Let us emphasize that for a given starting point and a distribution of cities, the route of the salesman in our model  is completely deterministic even though it resembles a random walk trajectory. Our model might be considered a version of the so-called deterministic random walk \cite{buni}.

\begin{figure}
\includegraphics[width=9cm]{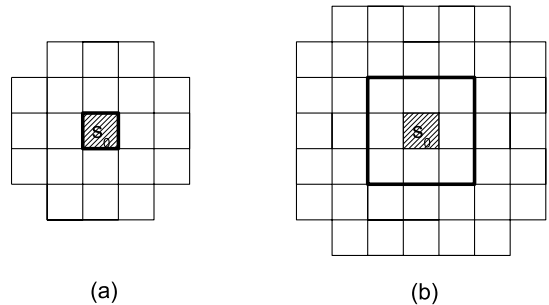}  \vspace{-1cm} 
\caption{ 
Sectorization of the two-dimensional space. For a trajectory that at a certain moment arrives at the sector $s_0$, the program searches for available cities within $s_0$. (i) If it finds some, then to find the nearest city it is sufficient to examine $s_0$ and its neighbouring sectors as indicated in~(a). This is because any city outside these sectors is further from any point in $s_0$ than the length of the diagonal of $s_0$ (which is the largest distance within $s_0$). (ii) If there are no available cities in $s_0$, the program searches for them in a larger area (the bold square in (b)). If it finds some, then to find the nearest city it is sufficient to examine sectors as indicated in (b). (iii) If no cities are found, the program examines even larger areas (until some cities are found).
\label{sectors}}
\end{figure}

To examine the asymptotic behavior of $\langle r^2\rangle$, it is essential to average over many configurations with large number of cities $N$ so that a relatively large part of the trajectory of the salesman was not affected by the boundaries. Because of these finite-size effects, we terminate the generation of trajectories after number of steps that is much smaller than~$N$. As a result, the $t$-th city on a trajectory is typically a close neighbour of the ($t$--1)-th city. Instead of searching for the nearest not yet visited city in the entire configuration, it would be much faster to restrict the search to the cities which are in the vicinity of the last visited city. Such a modification can be implemented by means of the sectorization of the space. In addition to coordinates, we store the information about a sector to which a given city belongs. Then, to find the $t$-th city, we search the sectors surrounding the \mbox{($t$--1)-th city} (see Fig.~\ref{sectors}). Let us emphasize that the above described sectorization does not introduce any approximation as for a given configuration of cities the algorithm generates exactly the same trajectory as the one without sectorization. In our simulations we used such mesh that each sector contained approximately 10 cities. (In the two-dimensional case, where the longest trajectories had to be generated, the sectorization reduced the simulation time by the factor $\sim 10^3$.)

\begin{figure}
\includegraphics[width=9cm]{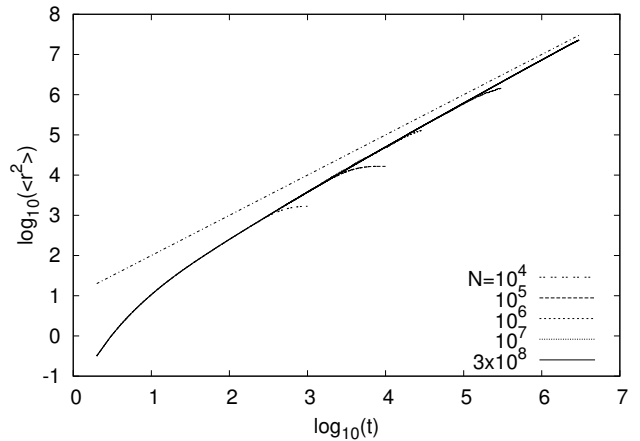}  \vspace{-1cm} 
\caption{ 
The logarithmic plot of the time dependence of $\langle r^2\rangle$ for several values of $N$ indicated in the right bottom. The straight line corresponds to the linear increase $\langle r^2\rangle \sim t$. For each~$N$, the plotted results are averages over at least $10^5$ independent configurations.
\label{walklog}}
\end{figure}
\section{Results}
Using the above algorithm, we simulated \mbox{$d$$=$2, 3, and 4} systems containing up to $3\times 10^8$ cities. Results for the two-dimensional case  are shown in Fig.~\ref{walklog}. Since the asymptotic slope seems to be bigger than unity, these data show a slightly superdiffusive behavior, but more quantitative analysis of the slope would be desirable. One should take into account that the asymptotic regime in the numerical data will be visible only in an intermediate interval since both the initial transit as well as large-$t$ data should be neglected (for long trajectories $\langle r^2\rangle$ is strongly affected by finite-size effects). To estimate $\alpha$, we first examine the time-dependent quantity $\alpha(t)$ defined as the local exponent estimated from data spanning over one decade and centered at $t$ (i.e., to estimate $\alpha(t)$ we used $\langle r^2\rangle$ calculated for $t'$ such that $\log_{10}(t)-0.5<\log_{10}(t')<\log_{10}(t)+0.5$).

\begin{figure}
\includegraphics[width=9cm]{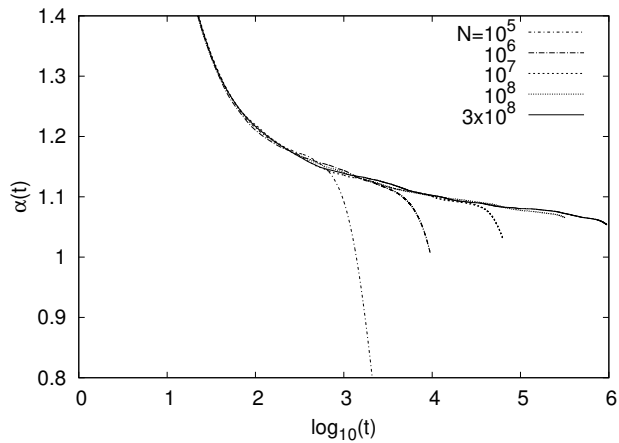}  \vspace{-1cm} 
\caption{ 
The time-dependent exponent $\alpha(t)$ calculated in the two-dimensional case for several values of~$N$.
\label{wykl2}}
\end{figure}
\begin{figure}
\includegraphics[width=9cm]{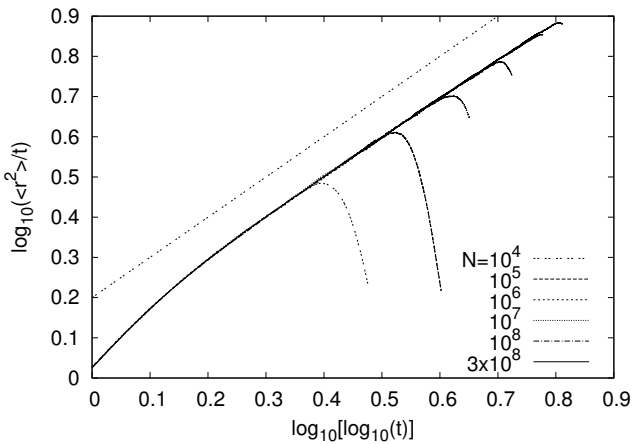}  \vspace{-1cm} 
\caption{ 
The plot of $\log_{10}(\langle r^2\rangle/t)$ vs. $\log_{10}[\log_{10}(t)]$ calculated in the two-dimensional  case. The dashed straight line has a unit slope.
\label{walkcorr}}
\end{figure}

The obtained results are presented in Fig.~\ref{wykl2}. As expected, for large $t$ all curves bent downward and in the middle range a plateau emerges. Our estimate based on  data around $t=10^5$ suggests that  $\alpha=1.08(1)$. However, such an estimation is very close to unity and in addition it seems to slowly decrease for increasing $t$. Such a slow decrease suggests that asymptotically $\alpha$ might be equal to unity but the power-law scaling is affected by logarithmic corrections. To examine such a possibility, we plotted $\log_{10}(\langle r^2\rangle/t)$ vs. $\log_{10}[\log_{10}(t)]$ (Fig.~\ref{walkcorr}). 
The asymptotic linearity of the data confirms the presence of the logarithmic corrections and, since the slope is close to unity, we obtain $\langle r^2\rangle \sim t\log_{10} (t)$.

\begin{figure}
\includegraphics[width=9cm]{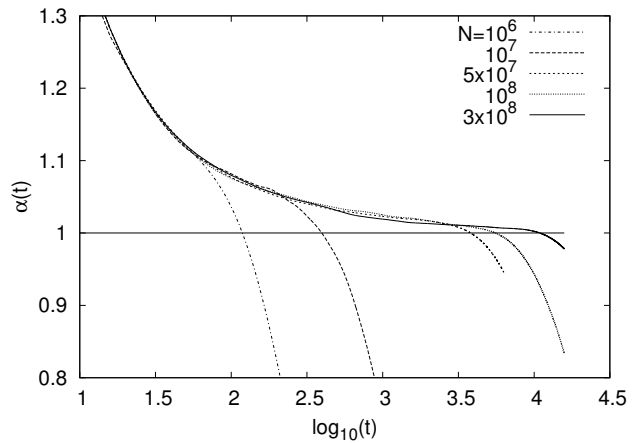}  \vspace{-1cm} 
\caption{ 
The time-dependent exponent $\alpha(t)$ calculated in the three-dimensional case for several values of~$N$.
\label{wykl3}}
\end{figure}

\begin{figure}
\includegraphics[width=9cm]{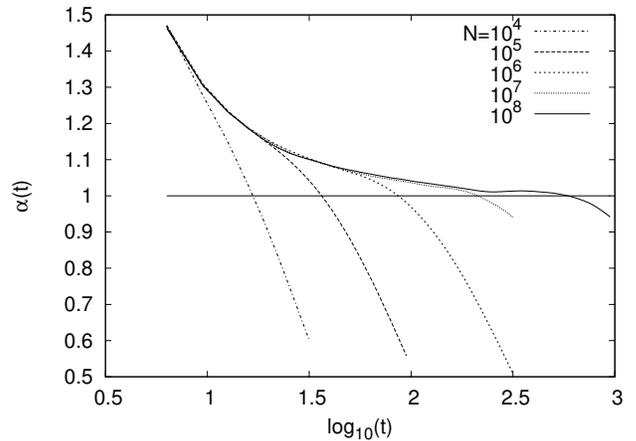}  \vspace{-1cm} 
\caption{ 
The time-dependent exponent $\alpha(t)$ calculated in the four-dimensional case for several values of~$N$.
\label{wykl4}}
\end{figure}

We used the same technique in the three- and four-dimensional versions, and the final results are shown in Fig.~\ref{wykl3} and Fig.~\ref{wykl4}, respectively. In these higher-dimensional cases, the linear size $L$ increases slower with the number of cities $N$, and that is why the curves bent downward for smaller values of $t$. The emerging plateaux indicate that $\alpha$ is very close to unity both for $d=3$ and~4.  Let us also notice that in the $d=1$ case the salesman, possibly after some initial switches, will stick to one side and then  move ballistically in the same direction, trivially implying $\alpha=2$.

\begin{figure}
\includegraphics[width=9cm]{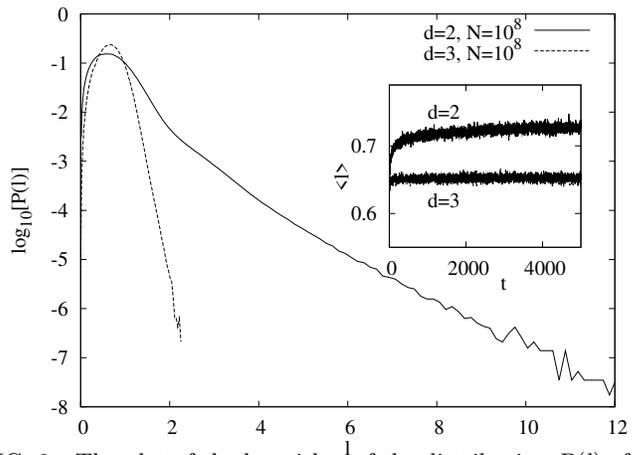}  \vspace{-1cm} 
\caption{ 
The plot of the logarithm of the distribution $P(l)$ of intervals $l$ calculated for paths of the length $t=5000$.  The inset shows that the variability of the average interval $\langle l \rangle$ (at time $t$) during such runs is rather small.
\label{distlog}}
\end{figure}

For the $d=2$ and 3 cases, we also analyzed the distributions of the length $l$ between successive cities on a trajectory. One should take into account, however, that these distributions are not entirely stationary and to some extent vary in time $t$ (in general, very long distances appear in the late-time evolution and they could be of the order of the system size~$L$ in the final stage). To reduce this variability, we  generated relatively short trajectories (most likely none of them reached the boundary of the system).

The logarithmic plot of these distributions (Fig.~\ref{distlog}) shows that, in the $d=2$ version, the distribution is much broader than in the $d=3$ version, and perhaps of a  qualitatively different large-$l$ tail. Let us notice that in both cases $\langle l \rangle$ is similar and of the order of unity (inset in Fig.~\ref{distlog}) but the recorded fluctuations of $l$ are several times larger in the $d=2$ case.
Let  us also observe that large-$l$ tails of these distributions do not seem to have a power-law form. This indicates that the logarithmic corrections seen  in the $d=2$ case are not due to sufficiently broad (and perhaps L\'evy-like) distributions of lengths $l$ but rather due to the development of long-time correlations. We will return to this point in the final part of the article.
\begin{figure}
\includegraphics[width=9cm]{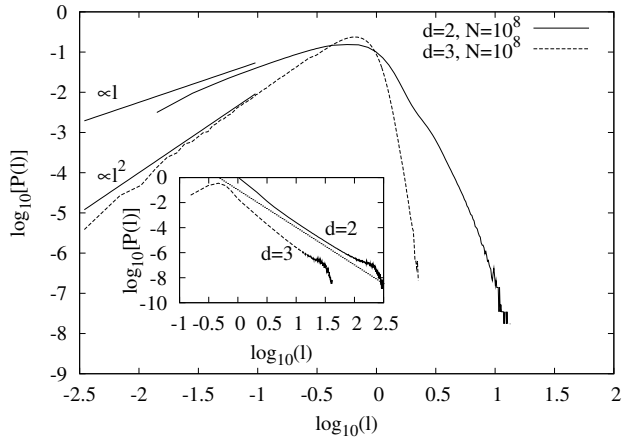}  \vspace{-0.5cm} 
\caption{ 
The same distribution $P(l)$ as in Fig.~\ref{distlog} but in the log-log scale. The inset shows the log-log plot of $P(l)$ for runs where all cities are visited  ($N=10^5$). The dotted line has the slope~-3 ($\sim l^{-3}$).
\label{distloglog}}
\end{figure}

The log-log plot of these distributions confirms the absence of power-law tails for large~$l$ (Fig.~\ref{distloglog}). It turns out that the other side of these distributions, namely the small-$l$ tails, can be understood with some simple arguments. Indeed, these tails do not  differ much from the small-$l$ tails in the initial distribution of cities. For simplicity, let us examine first the $d=1$ case, where $N$--1 cities are uniformly distributed on the interval of the length $L=N$ and one city is placed in the center of the interval (i.e., it has the coordinate $L/2$).  Calculating the probability density $P_1(x)$ that exactly one city is at the distance $x$ from the central city (to the right or left) and
$N$--2 cities are further than $x$ from it, in the limit $N\rightarrow\infty$, we obtain $P_1(x)=\frac{2(N-1)}{L}(1-\frac{2x}{L})^{N-2} = 2\exp(-2x)$. Generalizing the above calculations to the $d=2$  and $d=3$ cases, we obtain $P_2(x) \sim x\exp(-\pi x^2)$ and $P_3(x) \sim x^2\exp(-\frac{4}{3}\pi x^3)$, respectively. As can be seen in Fig.~\ref{distloglog}, the small-$l$ tails are indeed close to the tails of the initial distributions both for $d=2$ and 3. One can also notice that  for large $x$ the initial distributions $P_2(x)$ and $P_3(x)$ decay very fast (faster than exponentially) while the corresponding distributions $P(l)$ decay slower (see Fig.~\ref{distlog}). Thus, the traveling salesman generates paths where short distances appear with approximately the same distribution as in the initial distribution but long distances are generated much more frequently.

It is perhaps interesting, especially in the context of modeling animal search strategies, that the distributions $P(l)$ become much broader when calculations are made on paths that visit all the cities (inset in Fig.~\ref{distloglog}). This is because in the final part of the tour the salesman has to visit all the "leftovers" that are typically quite distant from each other. These numerical data suggest that in the two-dimensional data the tail of the resulting distribution might even have the power-law decay, $P(l)\sim l^{-\mu}$ and $\mu \sim 3$, and thus could be consistent with  the limiting L\'evy distribution. Let us emphasize, however, that, for paths that visit all the cities,  the generated sequence of lengths is nonstationary (early- and late-stage statistics  differ very much), and assembling the data into a single distribution is rather questionable.
Let us also note that there are some claims that the  L\'evy distributions in the foraging data appeared most likely due to a spurious inclusion of late-stage flights \cite{edwards}. These flights, e.g.,  of albatrosses returning to their nests, were obviously very long but should not be considered a part of the searching pattern.

\section{Final remarks}
In this section we briefly summarize the overall behavior of our model and comment on possible relations with some other statistical-mechanics models. The memory of the salesman restricts its movements and effectively pushes the salesman out of regions that have already been visited. For $d=1$, this strategy trivially leads to the superdiffusive behavior with $\alpha=2$. For $d=3$ and 4, the long-time memory turns out to be basically irrelevant and an  ordinary diffusion ($\alpha=1$) is restored. The  dimension $d=2$ seems to  be the critical dimension separating these two regimes and the logarithmic corrections are an expected feature in such a case.
Let us note that a similar memory mechanism  generates a superdiffusive behavior in the so-called self-avoiding walk problem \cite{sokal}. However, this model of a polymer is known to be superdiffusive for $d=1$ ($\alpha=2$), $d=2$ ($\alpha=1.5$), and $d=3$ ($\alpha=1.2$), with the normal diffusion resuming for $d>3$.  
Our model seems to be related to yet another model of the self-avoiding walk, introduced by Amit {\it et al.} \cite{amit}, where the walker tries to avoid lattice sites that it has already visited. Renormalization-group arguments show that for such a model the critical dimension equals to 2 and the logarithmic corrections are present \cite{amit,peliti}  in the critical dimension. It is likely that, at the coarse-grained level, our model is described by the same field theory as the model of Amit {\it et al.} Its direct applicability to search strategies is rather limited but our model might indicate that the very dimension of the search space plays an important role. In particular, it suggests that much broader distributions should appear in the two-dimensional rather than in the three-dimensional search spaces. 
Since a horizontal span of the search space for albatrosses or sharks is much larger than a  vertical one, perhaps this space could be considered as effectively two-dimensional and, as our work suggests, broad distributions might appear in such a case. Thus it would be interesting to examine species for which a search space is less anisotropic.
\begin{acknowledgments}
We acknowledge access to the computing
facilities at Pozna\'n Supercomputing and Networking Center.
\end{acknowledgments}

\end {document}